\documentclass[preprint,usenatbib,twocolumn]{aa}  
\usepackage{graphicx}
\usepackage[]{hyperref}

\usepackage{verbatim}
\usepackage{threeparttable}
\usepackage{subcaption}
\usepackage{enumitem}% http://ctan.org/pkg/enumitem
\setlist[itemize]{noitemsep, topsep=0pt}
\usepackage{tabu}
\usepackage{gensymb}
\usepackage{txfonts}
\usepackage{makecell}
\usepackage{afterpage}
\usepackage{multirow}
\usepackage[normalem]{ulem}
\usepackage{subcaption}
\defcitealias{Turner2021}{T21}

\usepackage{color}

\usepackage{afterpage}
\defcitealias{Turner2019}{T19}

\begin{document} 
\title{Follow-up LOFAR observations of the $\tau$~Bo\"{o}tis exoplanetary system}
   \author{Jake D. Turner\inst{1,2},
           Jean-Mathias Grie{\ss}meier\inst{3,4},
           Philippe Zarka\inst{4,5},
           Xiang Zhang\inst{5}, Emilie Mauduit\inst{5}
          }
   \institute{Department of Astronomy and Carl Sagan Institute, Cornell University, Ithaca, NY, USA\\
              \email{jaketurner@cornell.edu}
     \and NHFP Sagan Fellow 
          \and Laboratoire de Physique et Chimie de l'Environnement et de l’Espace (LPC2E) Universit\'{e} d’Orl\'{e}ans/CNRS, Orl\'{e}ans, France
    \and  Observatoire Radioastronomique de Nan\c{c}ay (ORN), Observatoire de Paris, PSL Research University, CNRS, Univ. Orl\'{e}ans, OSUC, 18330 Nan\c{c}ay, France
     \and LESIA, Observatoire de Paris, CNRS, PSL, Meudon, France
     }
   \date{}

\abstract
  % context heading (optional)
  % {} leave it empty if necessary  
   { 
Observing the radio emission from exoplanets is among the most promising methods to detect their magnetic fields and a measurement of an exoplanetary magnetic field will help constrain the planet's interior structure, star-planet interactions, atmospheric escape and dynamics, and habitability. Recently, circularly polarized bursty and slow emission from the $\tau$~Bo\"{o}tis ($\tau$~Boo) exoplanetary system was tentatively detected using LOFAR (LOW-Frequency ARray) beamformed observations. If confirmed, this detection will be a major contribution to exoplanet science. However, follow-up observations are required to confirm this detection. 
   } 
  % aims heading (mandatory)
   {
   Here, we present such follow-up observations of the $\tau$~Boo system using LOFAR. These observations cover 70$\%$ of the orbital period of $\tau$~Boo~b including the orbital phases of the previous tentative detections. 
   }
  % methods heading (mandatory)
   {
   We used the \texttt{BOREALIS} pipeline to mitigate radio frequency interference and to search for bursty and slowly varying radio signals. \texttt{BOREALIS} was previously used to find the tentative radio signals from $\tau$~Boo. 
   }
  % results heading (mandatory)
   {
  Our new observations do not show any signs of bursty or slow emission from the $\tau$~Bo\"{o}tis exoplanetary system.}
  % conclusions heading (optional), leave it empty if necessary 
  {
   The cause for our non-detection is currently degenerate. It is possible that the tentative radio signals were an unknown instrumental systematic or that we are observing variability in the planetary radio emission due to changes in its host star. More radio data (preferably multi-site) and ancillary observations (e.g. magnetic maps) are required to further investigate the potential radio emission from the $\tau$~Bo\"{o}tis exoplanetary system.  
 }

%% Keywords should appear after the \end{abstract} command. 
%% See the online documentation for the full list of available subject
%% keywords and the rules for their use.

   \titlerunning{Follow-up LOFAR observations of $\tau$ Bootis}
   \authorrunning{J.D. Turner, P. Zarka, J.-M. Grie{\ss}meier, et al} 
   \maketitle

%\maketitle

%-------------------------------------------------------------------

\section{Introduction}

A confirmed direct detection of exoplanetary radio emission has been elusive for decades. Observing the radio emission from exoplanets is among the most promising methods to detect their magnetic fields (\citealt{G2015}) and a measurement of an exoplanetary magnetic field will help constrain the planet's interior structure, star-planet interactions, and atmospheric escape (e.g. \citealt{Zarka2015SKA,G2015,Lazio2016,Griessmeier17PREVIII,Lazio2018,Griessmeier2018haex,Zarka2018haex,Lazio2019}). A magnetic field measurement can help also help break degeneracies in the mass-radius diagram \citep{Lazio2019}. Additionally, atmospheric dynamics may be influenced by the presence of a planetary magnetic field (e.g. \citealt{Perna2010a,Rauscher2013,Rogers2014b, Hindle2021}) and Ohmic dissipation might be one of the factors causing the anomalously large radii of hot Jupiters (e.g. \citealt{Perna2010a,Perna2010b,Knierim2022}). Finally, a magnetic field might be one of the many properties needed on Earth-like exoplanets to sustain their habitability 
(e.g. \citealt{Gr2005,Gr2015_cosmic,Gr2016,Lammer2009,Kasting2010,Owen2014,Lazio2010a,Lazio2016,Meadows2018haex,McIntyre2019,Gronoff2020,Green2021}).

Observations of auroral radio emissions produced via the Cyclotron Maser Instability (CMI) mechanism (\citealt{Wu1979,Zarka1998,Treumann2006}) have been used to directly measure the magnetic field strengths of some of the magnetized planets in our Solar System (\citealt{Burke1955,Zarka1998}). CMI radio signals are emitted at the local electron cyclotron frequency in the source region. Hence, the maximum gyrofrequency is directly proportional to the maximum magnetic field near the planetary surface. CMI radio emission is circularly polarized, beamed, and time-variable (e.g. \citealt{Zarka1998,Zarka2004}).

Many theoretical and observational studies have focused on studying exoplanetary radio emissions over the past few decades. The reviews by \citet{Zarka2015SKA} and \citet{G2015,Griessmeier17PREVIII} cover the subject in great detail. A large body of theoretical work\footnote{e.g. \citealt{Pater2000pras,Farrell2004,Lazio2004,Stevens2005,Griessmeier05AA,Griessmeier07PSS,Griessmeier2007_AA,Jardine2008,Vidotto2010r,Hess2011,Nichols2011,Nichols2012,Vidotto2012,Saur2013,See2015,Vidotto2015,Nichols2016,Fujii2016,Vidotto2017,Weber2017,Weber2017pre8,Weber2018,Lynch2018,Zarka2018haex}; \citealt{Wang2019}; 
\citealt{Kavanagh2019,Vidotto2019};
\hspace{5em} 
\citealt{Shiohira2020, Gasperin2020,Green2021};
\citealt{Sciola2021,Narang2021,Yasunori2021,Cendes2022}
} has been published since the foundational theoretical studies of \citet{Zarka1997pre4}, \citet{Farrell1999}, \citet{Zarka2001}, and \citet{Zarka2007}. Some of these studies aim to develop theoretical models that can predict the frequency and intensity of radio emissions observed from Earth. A variety of scaling laws have been developed to estimate the magnetic field strengths of exoplanets (e.g. \citealt{Farrell1999,SL2004,Griessmeier2007_AA,Christensen2009,Christensen2010,Reiners2010}). Likewise, the Radio-Magnetic Bode's law is a common approach for estimating the radio emission intensity of these planets (e.g. \citealt{Farrell1999,Lazio2004,Zarka2007,Griessmeier2007_AA,Griessmeier17PREVIII,Zarka2018,Zarka2018haex}).

Over the past few decades, there have been many exoplanet radio observing campaigns, but none of these observations have resulted in an unambiguous detection\footnote{e.g. \citealt{Yantis1977,Winglee1986,Zarka1997pre4,Bastian2000,Farrell2003,Ryabov2004,Shiratori2006,George2007,Lazio2007,Smith2009,Lecavelier2009,Lecavelier2011,Lazio2010a,Lazio2010b,Stroe2012,Hallinan2013,Murphy2015,Lynch2017,Turner2017pre8,Gorman2018,Lenc2018,Lynch2018,Route2019,Narang2021,Gasperin2020,Green2020,Turner2021,Narang2021,Cendes2022,Narang2022,Narang2023,Narang2023b,Route2023,Turner2023,Shiohira2024,Narang2024}}. There have been a few tentative detections \citep{Lecavelier2013,Sirothia2014,Vasylieva2015,Bastian2018}, but none of them have been confirmed by follow-up observations. The many degenerate reasons for the non-detections are discussed in great detail in the review articles \citet{Zarka2015SKA} and \citet{G2015,Griessmeier17PREVIII}. Some reasons for the non-detections could be that the observations were not sensitive enough (e.g. \citealt{Bastian2000}), there was no emission at the observed frequencies due to a low magnetic field strength (e.g. \citealt{Hallinan2013}), Earth was outside the CMI beaming pattern (\citealt{Hess2011,Ashtari2022}), and/or the CMI mechanism was not operating (e.g. plasma frequency of the planet's ionosphere was greater than the cyclotron frequency; \citealt{Griessmeier2007_AA,Weber2017,Weber2017pre8,Lamy2018,Weber2018}). By contrast, advancement has been made in detecting free-floating planets near the brown dwarf boundary (\citealt{Kao2016,Kao2018}) and stellar emission potentially created by star-planet interactions (e.g. \citealt{Vedantham2020,Callingham2021,Perez2021,Pineda2023,Trigilio2023,BlancoPozo2023}).

\begin{figure}[!htb]
\centering
\includegraphics[width=0.5\textwidth]{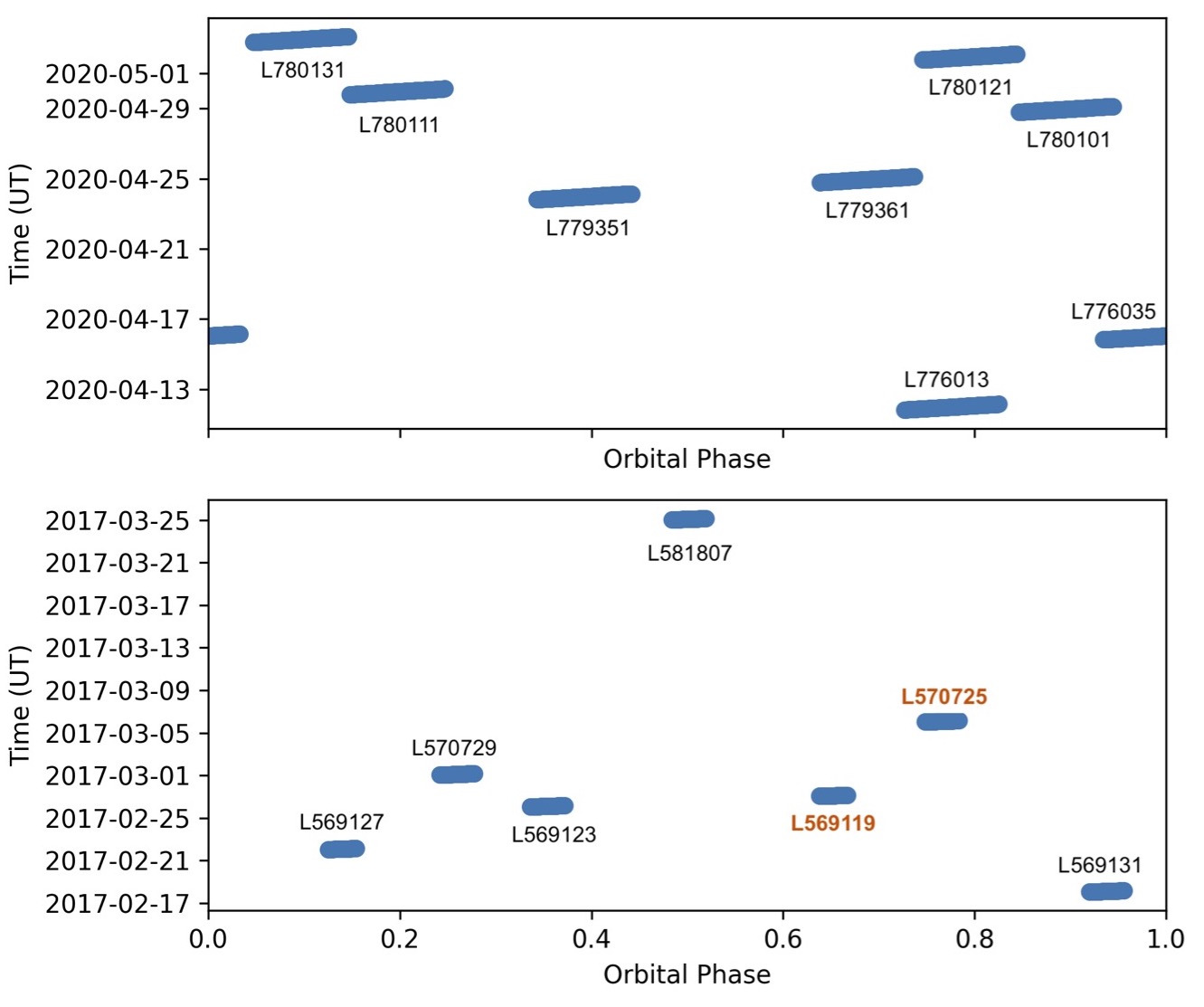}
\caption{Orbital phase coverage for the LOFAR observations of $\tau$~Boo~b. The observations from this study and \citet{Turner2021} are shown in the top and bottom panels, respectively. The orbital phase was calculated relative to periastron of $\tau$~Boo~b using the ephemeris (Period = 3.312463 days, T(0)$_{per}$=2446957.81 JD) from \citet{Wang2011}. The $\tau$~Boo LOFAR observations with the tentative detections (L569119 and L570725) from \citet{Turner2021} are displayed in red. }
\label{fig:orbits}
\end{figure}

In a recent study, circularly polarized bursty and slow emission from the $\tau$~Bo\"{o}tis ($\tau$~Boo) exoplanetary system was tentatively detected by \citet[hereafter T21]{Turner2021} using LOFAR (LOw Frequency ARray; \citealt{vanHaarlem2013}) beamformed observations taken in 2017. The slow and burst emission were seen at orbital phases of 0.65 and 0.8, respectively, in relation to the $\tau$~Boo b's periastron (see Figure \ref{fig:orbits}). The bursty emission was seen from 15--21 MHz, while the slower emission was present in the 21--30 MHz range. \citetalias{Turner2021} did not identify any plausible sources of false positives for the bursty emission, but when it came to the slower emission, uncertainty arose due to the possibility of unknown systematic effects causing the signal. Assuming an astronomical origin for the signal, \citetalias{Turner2021} concluded that the signals originated from the $\tau$~Boo system, with the most likely cause being CMI radio emission from the exoplanet $\tau$ Boo b. Under the premise of a planetary origin for the radio signals, \citetalias{Turner2021} derived planetary magnetic field constraints that are consistent with theoretical predictions (\citealt{Griessmeier2007_AA,Griessmeier17PREVIII}). More recent calculations (\citealt{Mauduit23PRE9,Ashtari2022}) using an updated version of these predictions also yielded magnetic moment and flux density values that aligned with the earlier theoretical estimates, in addition to matching the observed phases and the handedness of the circular polarization (\citealt{Ashtari2022}). Follow-up observations were highly advocated by \citetalias{Turner2021} to confirm their tentative detections and to search for periodicity in the signal. 

Motivated by the tentative detection of radio emission from $\tau$~Boo, we performed a large follow-up campaign in 2020. This campaign was coordinated between four different radio telescopes including LOFAR and NenuFAR (New Extension in Nan\c{c}ay Upgrading LOFAR; \citealt{Zarka2020}). In this paper, we only present the LOFAR data of $\tau$~Boo since the sensitivity of LOFAR has been well characterized using simulated Jupiter radio emission data (\citealt[hereafter T19]{Turner2019}) and a complete data reduction pipeline is available (\citetalias{Turner2019,Turner2021}). A preliminary analysis of the NenuFAR data is presented elsewhere \citep{Turner2023}.

In this paper, Sections \ref{sec:obs}, \ref{sec:pipeline}, and \ref{sec:dataAnalysis} describe the $\tau$~Boo observations, data processing, and data analysis, respectively. Section \ref{sec:discussion} discusses the possible reasons for our non-detections and future steps.

%%%%%%%%%%%%%%%%%%%%%%%%%%%%%%%%%%%%%%%%%%%%%%%%%%%%%%%%%%%%%
%%%%%%%%            Observations                       %%%%%%
%%%%%%%%%%%%%%%%%%%%%%%%%%%%%%%%%%%%%%%%%%%%%%%%%%%%%%%%%%%%%
\section{Observations} \label{sec:obs}
\begin{table}[!tbh]
%\begin{threeparttable}
\centering
\caption{Summary of the observations of $\tau$~Bo\"{o}tis in LOFAR Cycle 13.
} 
\begin{tabular}{ccccc}
\hline 
\hline
Obs $\#$  &  LOFAR ID  &  Date $\&$ Start  & $\Delta t$& Phase    \\
          &            & Time (UTC)          & (hrs)       &      \\
\hline

1	&	L776013	&	2020-04-11 20:00   & 8             & 0.73--0.83 \\
2	&	L776035	&	2020-04-15	20:00& 8               & 0.93--0.03 \\
3	&	L779351	&	2020-04-23	19:30 & 8              & 0.34--0.44\\
4	&	L779361	&	2020-04-24	19:00 & 8              & 0.64--0.74  \\
5	&	L780101	&	2020-04-28	19:00 & 8              & 0.85--0.95  \\
6	&	L780111$^{a}$	&	2020-04-29	19:00 & 8      & 0.15--0.25   \\
7	&	L780121	&	2020-05-01	18:30 & 8              &  0.75--0.85  \\
8	&	L780131	&	2020-05-02	18:30 & 8              & 0.05--0.15\\
 
\hline
                \multicolumn{5}{c}{B0809+74 [82 min.]} \\  
\hline
1	&	L776009	&	2020-04-11 1949	&	0.17	& -- \\
2	&	L776031	&	2020-04-15 1949	&	0.17	& --\\
3	&	L779347	&	2020-04-23 1919	&	0.17	& --\\
4	&	L779357	&	2020-04-24 1849	&	0.17	& --\\
5	&	L780097	&	2020-04-28 1849	&	0.17	& --\\
6	&	L780107	&	2020-04-29 1849	&	0.17	& --\\
7	&	L780117	&	2020-05-01 1819 &	0.17	& --\\
8	&	L780127	&	2020-05-02 1819 &	0.17	& --\\
\hline
\hline
\end{tabular}
\label{tb:obs}
 \tablefoot{
Column 1: observation number. Column 2: LOFAR observation ID. Column 3: date and start time of the observation (UTC). Column 4: duration ($\Delta$t) of the observation. 
Column 5: $\tau$~Boo~b orbital phases of the observation relative to periastron and using the ephemeris (Period = 3.312463 days, T(0)$_{per}$=2446957.81 JD) from \citealt{Wang2011}.

$^{a}$For L780111, only the last 3 hours of this observation are useful due to strong RFI at the start of the observations. 
}
%\end{threeparttable}
\end{table}

%%%%%%%%%%%%%%%%%%%%%%%%%%%%
% Setup Table 
%%%%%%%%%%%%%%%%%%%%%%%%%%%%%

\begin{table}[!tb]
\centering
\caption{Setup of the LOFAR-LBA beamformed observations of $\tau$~Bo\"{o}tis}
\begin{tabular}{ccc}
\hline 
\hline
Parameter   & Value  & Units   \\ 
\hline
\hline 
Array Setup                     & LOFAR Core                &     \\
Number of Stations              & 24                        &\\
Beams                           & ON $\&$ 3 OFF                & \\ 
Configuration                   & LBA outer antennas        &    \\
Antennas per Station            & 48                        &            \\
Minimum Frequency               & 15                      & MHz   \\
Maximum Frequency               & 39                    & MHz   \\
Subbands Recorded               & 244                        &    \\
Subband Width                    & 195                       &kHz \\
Channels per Subband            & 64                        &   \\
Frequency Resolution  ($b$)     & 3.05                      & kHz   \\
Time Resolution ($\tau$)        & 10.5                      & msec  \\
Beam Diameter\tnote{a}          & 13.8                      & arcmin    \\
Stokes Parameters               & IQUV                      &   \\
\hline
\end{tabular}
    \tablefoot{
  \tablefoottext{a}{Calculated at 30 MHz (\citealt{vanHaarlem2013})}.}
\label{tb:setup}
\end{table}

All of the observations presented here were taken with the Low Band Antenna (LBA) of LOFAR (\citealt{vanHaarlem2013}) using beamformed mode (\citealt{Stappers2011}). We obtained 8 exoplanet observations (each 8 hours long) for a total of 64 hours, plus 8 observations of the pulsar B0809+74 totaling 82 minutes.  Similar to our previous studies (\citealt{Turner2017pre8}, \citetalias{Turner2019}, \citetalias{Turner2021}, \citealt{Turner2023}), the pulsar observations are used as a calibrator to study systematics in the data and to ensure the reliability of the processing. An example OFF-beam dynamic spectrum is shown in Figure \ref{fig:Dynspec_OFF}. The dates and times of the observations can be found in Table \ref{tb:obs}. The setup of the observations can be found in Table \ref{tb:setup}; it is similar to that in \citetalias{Turner2021}. Most importantly, we ensure we cover the 15--30 MHz range where \citetalias{Turner2021} detected their tentative signals. We only present the Stokes-V\footnote{Similar to \citetalias{Turner2021}, we only analyze $V^{'}$ as defined in equations 9--11 of Section 4.2.1 in \citetalias{Turner2019}.} in this paper since the tentative detections were seen in Stokes-V (\citetalias{Turner2021}).

In this study, we compare our on-target beam (``ON-beam'') to several simultaneous beams pointing to nearby locations in the sky (``OFF-beam 1'', ``OFF-beam 2'' and ``OFF-beam 3''). This is similar to our previous studies \citep{Turner2017pre8,Turner2019,Turner2021} except we now use three OFF-beams instead of two to further investigate systematic effects in the analysis. This method relies on the assumption that the OFF beams can effectively characterize and account for ionospheric fluctuations, radio frequency interference, and any other systematic errors that may be present. The beamformed LOFAR observations of 55 Cancri, $\upsilon$~Andromedae, and $\tau$~Boo demonstrated that this assumption holds true (\citetalias{Turner2021}). 

Our new LOFAR observations cover 70$\%$ of the orbit of $\tau$~Boo~b as seen in Figure \ref{fig:orbits}. Due to the emission geometry (\citealt{Hess2011,Ashtari2022}), it is anticipated that the emission will only be directed towards Earth for a limited period. Therefore, the observation windows were selected to ensure the widest possible orbital coverage to maximize the chances to detect beamed emission. Most importantly, we also obtained two new observations around phase 0.65 (2020-04-11 and 2020-05-01) and one new observation around phase 0.8 (2020-04-15) that cover the same phases as the tentative slow emission and burst emission signal from \citetalias{Turner2021}, respectively. These phases are most likely to have detectable emission as shown in the modeling of the time-dependent CMI beaming effect by \citet{Ashtari2022}.

%----------------------------------------------------------------------------------
\section{Data pipeline } \label{sec:pipeline}

\begin{figure}[!thb]
\centering  
  \vspace{-0.5em}
  \begin{tabular}{c}
  \vspace{-1.5em}
   \begin{subfigure}[c]{0.47\textwidth}
        \centering
        \caption{}
 \includegraphics[width=\textwidth]{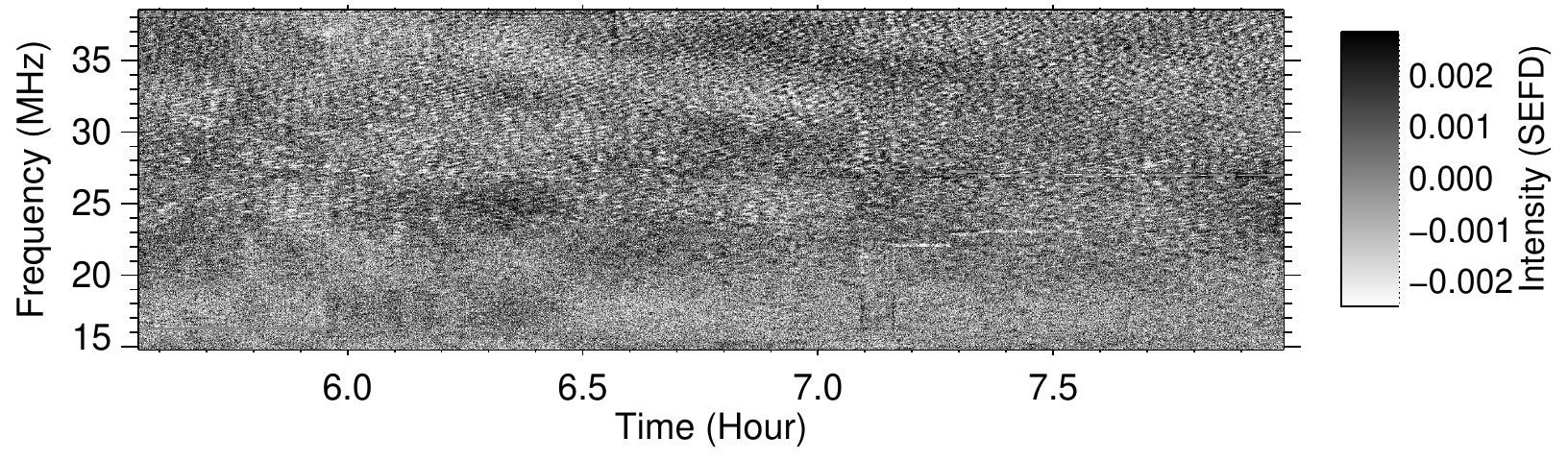}
        \label{}
    \end{subfigure} \\
       \begin{subfigure}[c]{0.47\textwidth}
        \centering
        \caption{}
 \includegraphics[width=\textwidth]{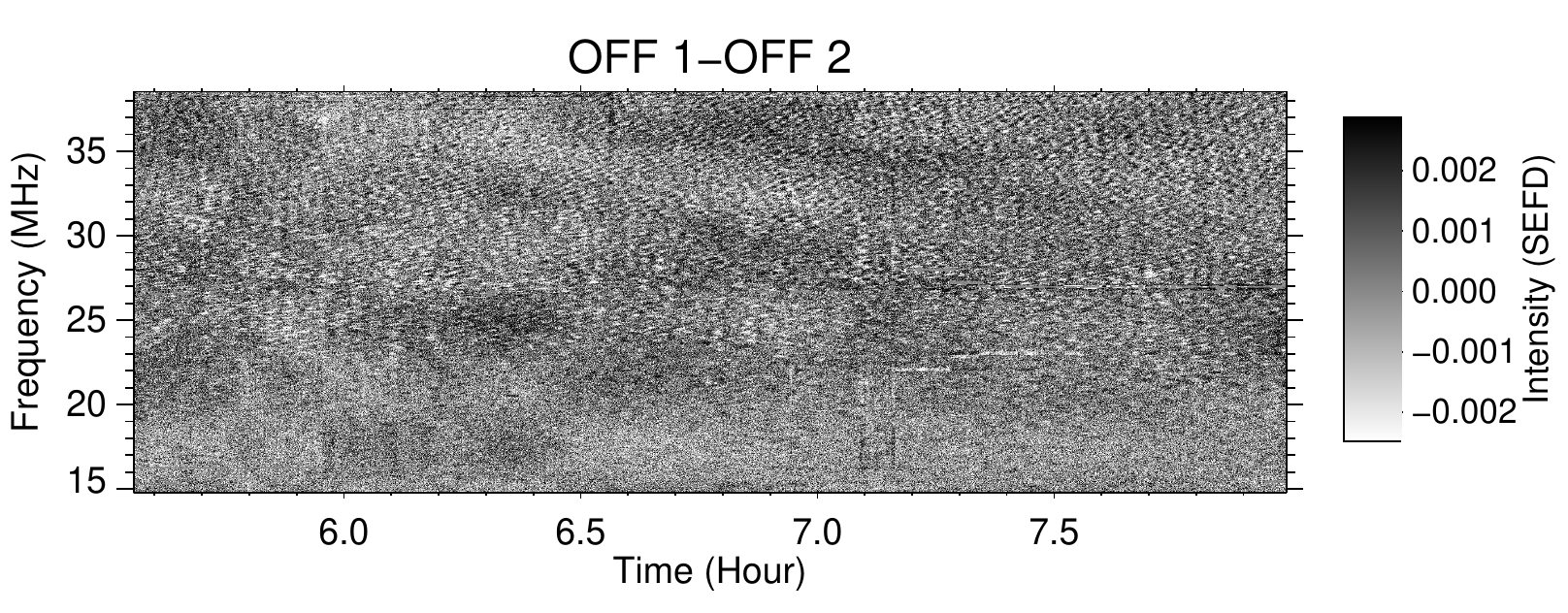}
        \label{}
    \end{subfigure}%
\end{tabular}
\vspace{-2em}
  \caption{Example dynamic spectrum of the OFF-beam (panel a) and the difference (panel b) between two OFF-beams in Stokes-V for the L780111 (2020-04-29) $\tau$~Boo observation. This dynamic spectrum has been processed with \texttt{BOREALIS}. } 
  \label{fig:Dynspec_OFF}
\end{figure}

All observations were processed with the \texttt{BOREALIS} (Beam-fOrmed Radio Emission AnaLysIS) pipeline (\citealt{Vasylieva2015,Turner2017pre8,Turner2019,Turner2021}. BOREALIS is split into two main components: processing and post-processing. The processing part of \texttt{BOREALIS} performs RFI mitigation, corrects for any large-scale time-frequency (t-f) response variations of the telescope (i.e. gain variations), and rebins the RFI mitigated and corrected data. All parts of the analysis were done exactly as in \citetalias{Turner2021}. The RFI mitigation combines four different techniques \citep{Offringa2010,Offringa2012AA,Offringa2012PhD,Zakharenko2013,Vasylieva2015} for optimal efficiency and processing time. The t-f response of the telescope is empirically determined assuming a quadratic dependence in time for each frequency. BOREALIS has been extensively tested on pulsar \citep{Turner2017pre8} and scaled Jupiter radio emission data \citep{Turner2019} to ensure optimal RFI mitigation and t-f response variation corrections without sacrificing detection capabilities. See Figure \ref{fig:Dynspec_OFF} for example OFF-beam dynamic spectrum processed through \texttt{BOREALIS}.

\begin{figure*}[!tbh]
\vspace{-0.5em}
    \centering 
    \begin{tabular}{cc}
    \vspace{-1.5em}
      \begin{subfigure}[c]{0.45\textwidth}
        \centering
        \caption{}
           \includegraphics[page=1,width=\textwidth]{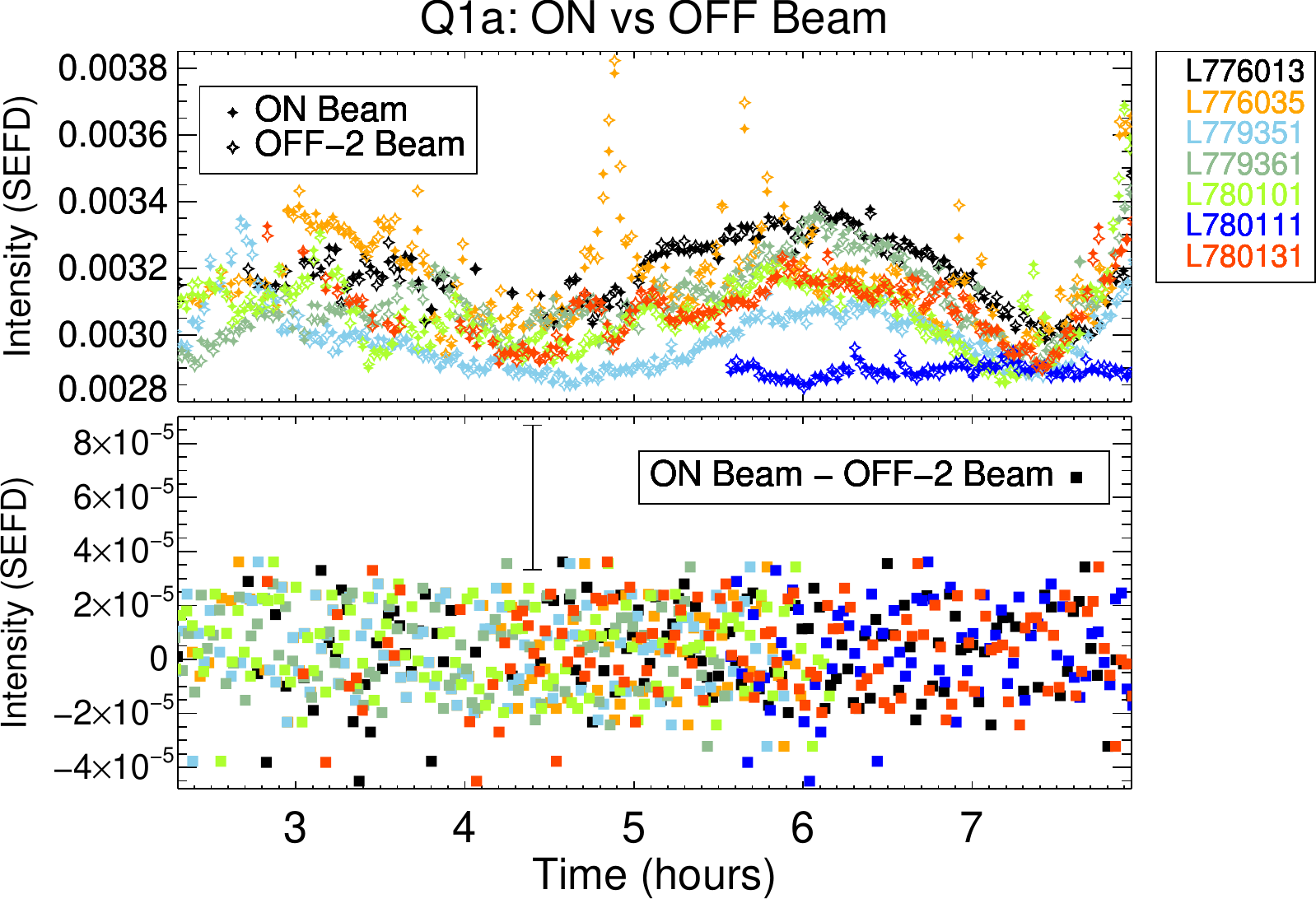}        
            \label{}
    \end{subfigure}%
    & 
         \begin{subfigure}[c]{0.45\textwidth}
        \centering
        \caption{}
        \includegraphics[page=1,width=\textwidth]{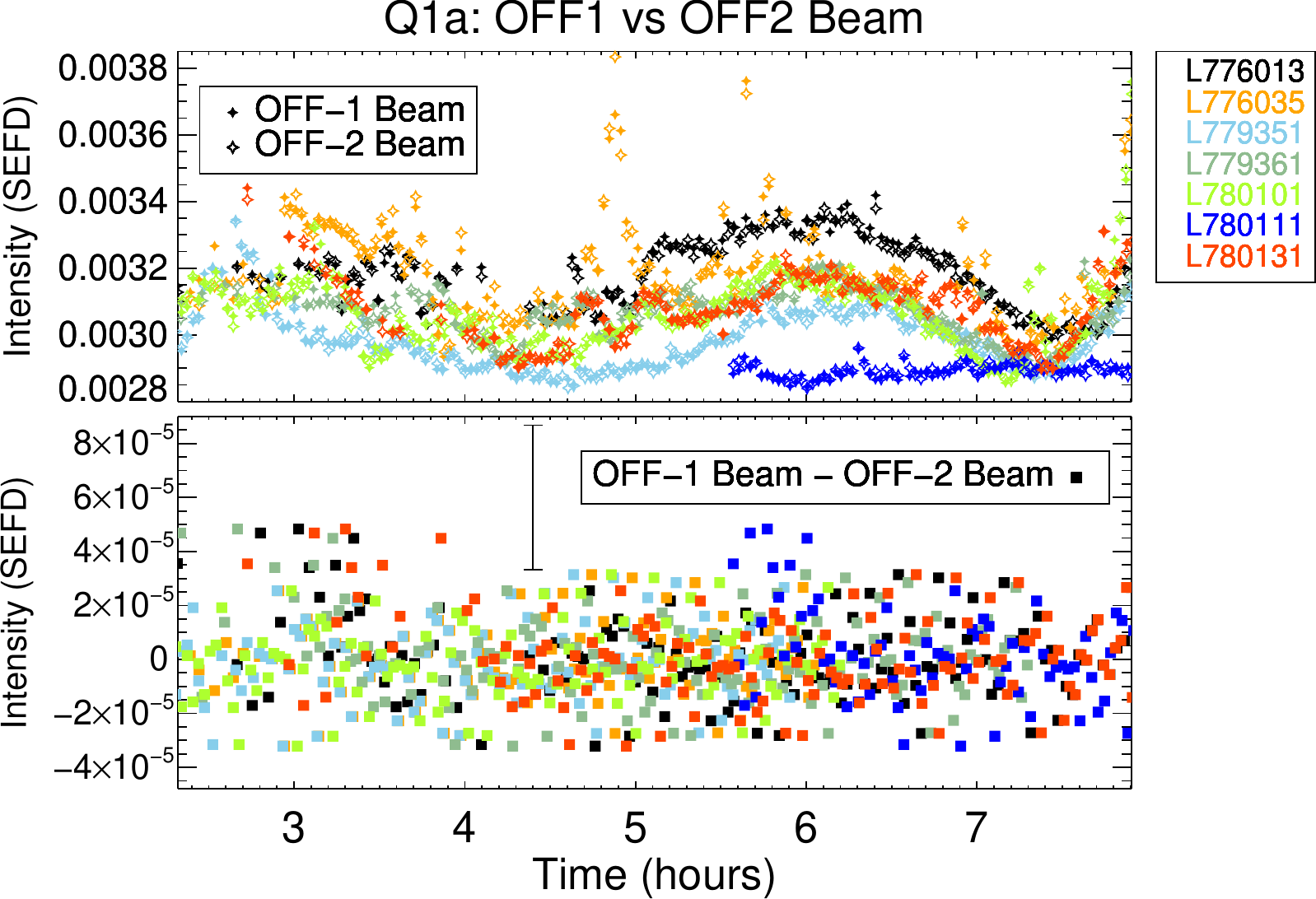}
        \label{}
    \end{subfigure}%
    \\
        \begin{subfigure}[c]{0.45\textwidth}
        \centering
        \caption{}
            \includegraphics[page=1,width=\textwidth]{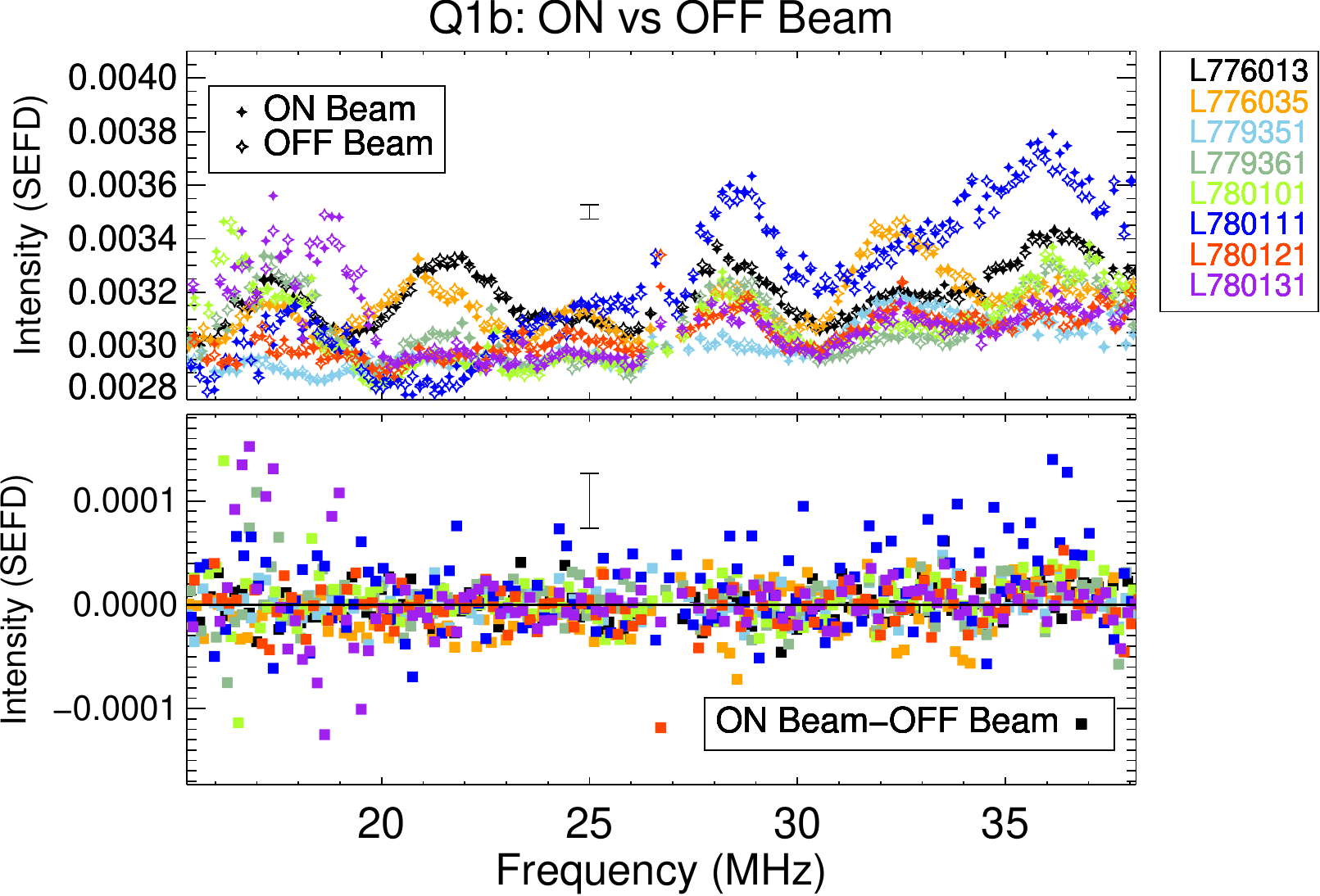}
            \label{}
    \end{subfigure}%
    & 
         \begin{subfigure}[c]{0.45\textwidth}
        \centering
        \caption{}
      \includegraphics[page=1,width=\textwidth]{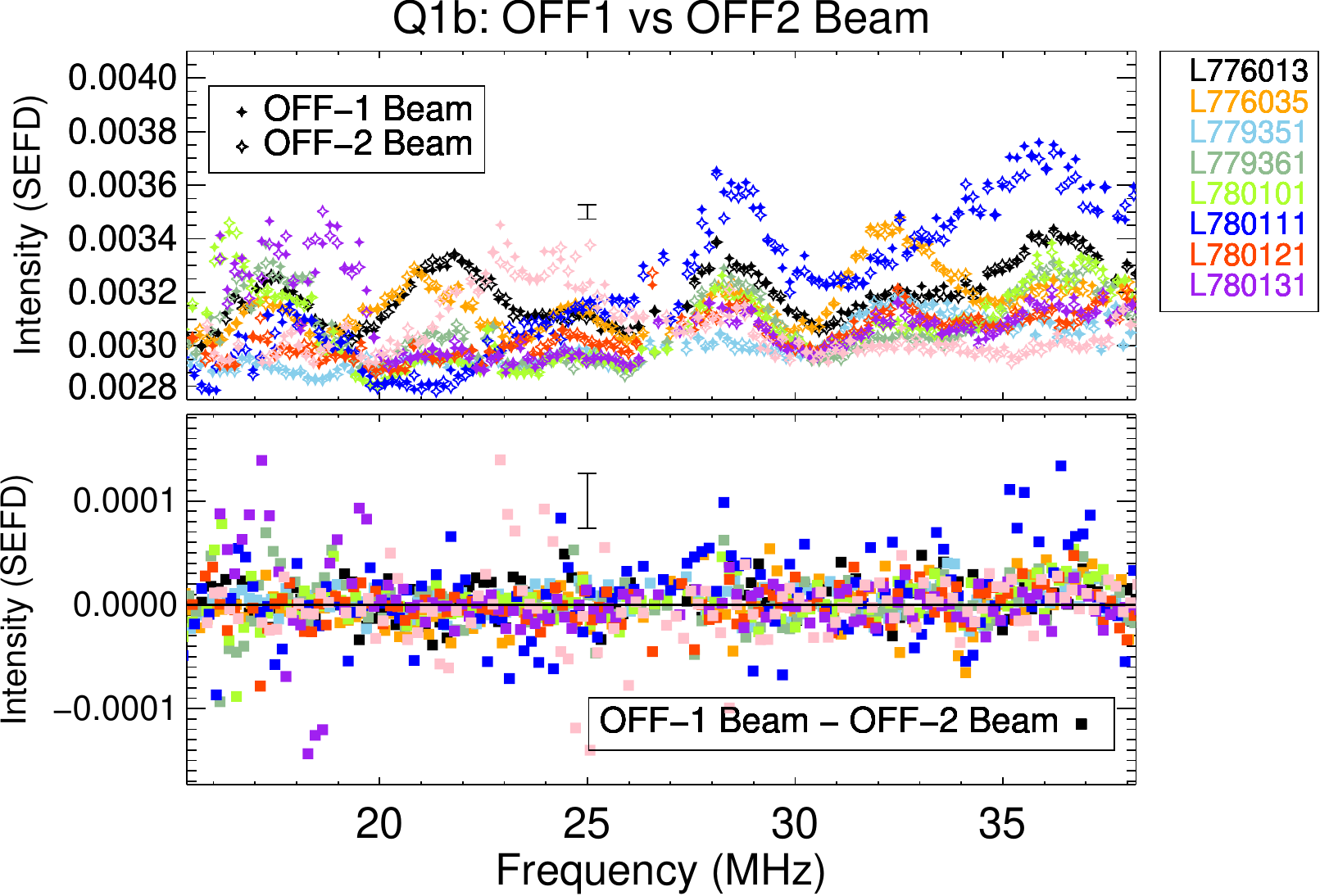}

        \label{}
    \end{subfigure}%
    \\ 
  %\begin{flushleft} \textbf{(a)} \end{flushleft} &  
  %\begin{flushleft} \textbf{(b)} \end{flushleft} \\
     %    \includegraphics[page=1,width=0.48\textwidth]{TauBoo_ON_overplot_Q1a.pdf} &   \includegraphics[page=1,width=0.48\textwidth]{TauBoo_OFF_overplot_Q1a.pdf}\\
        %   \begin{flushleft} \textbf{(c)} \end{flushleft} &  \begin{flushleft} \textbf{(d)} \end{flushleft} \\
         %\includegraphics[page=1,width=0.48\textwidth]{TauBoo_ON_overplot_Q1b.pdf} & \includegraphics[page=1,width=0.48\textwidth]{TauBoo_OFF_overplot_Q1b.pdf}
    \end{tabular}
    \vspace{-1em}
    \caption{Time series (Q1a: panel a and b) and  Integrated spectrum (Q1b: panel c and d) for different beams (ON, OFF-1, OFF-2) for the follow-up LOFAR $\tau$~Bo\"{o}tis observations. The two-sigma error bars are shown in the black brackets. Large-scale features can be seen for all dates, however, they change between observations.  No distinct emission in the ON-beam is seen when compared to the OFF-beams. There are no large-scale differences seen between the three beams for any date. The third OFF-beam is indistinguishable from the other OFF-beams within the noise.    }
    \label{fig:Q1_all}
\end{figure*}

\begin{figure*}[p]
\centering
  \begin{tabular}{cc}
    \begin{subfigure}[c]{0.45\textwidth}
        \centering
        \caption{}
  \includegraphics[width=\textwidth,page=1]{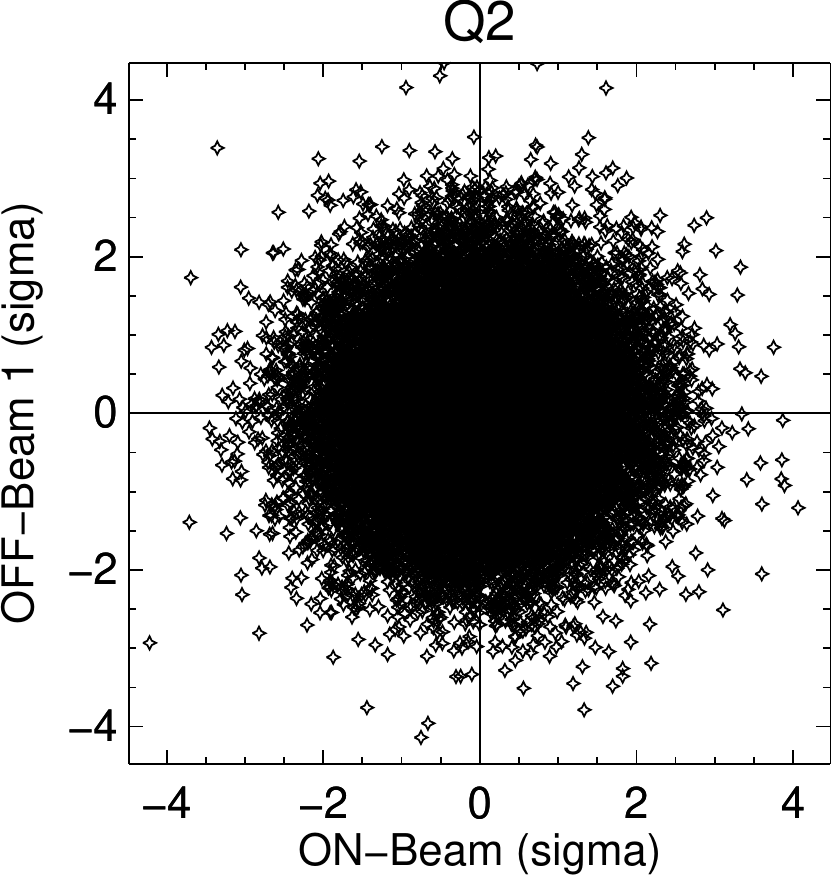} 
        \label{}
    \end{subfigure}%
    & 
         \begin{subfigure}[c]{0.45\textwidth}
        \centering
        \caption{}
 \includegraphics[width=\textwidth,page=2]{L776035_14.6to26.7MHz_Nondetection-crop.pdf}

        \label{}
    \end{subfigure}%
    \\
    \begin{subfigure}[c]{0.45\textwidth}
        \centering
        \caption{}
  \includegraphics[width=\textwidth,page=4]{L776035_14.6to26.7MHz_Nondetection-crop.pdf}
        \label{}
    \end{subfigure}%
    & 
         \begin{subfigure}[c]{0.45\textwidth}
        \centering
        \caption{}
   \includegraphics[width=\textwidth,page=6]{L776035_14.6to26.7MHz_Nondetection-crop.pdf}
        \label{}
    \end{subfigure}%
    \\
    
        \begin{subfigure}[c]{0.45\textwidth}
        \centering
        \caption{}
 \includegraphics[width=\textwidth,page=8]{L776035_14.6to26.7MHz_Nondetection-crop.pdf}
        \label{}
    \end{subfigure}%
    & 
         \begin{subfigure}[c]{0.45\textwidth}
        \centering
        \caption{}
   \includegraphics[width=\textwidth,page=10]{L776035_14.6to26.7MHz_Nondetection-crop.pdf} 
        \label{}
    \end{subfigure}%
    \\
\end{tabular}
\vspace{-2em}
  \caption{Non-detection of burst emission for $\tau$~Boo in the L776035 (2020-04-15) observation between 15-27 MHz in Stokes-V ($|V^{'}|$). 
  \textit{Panel a:} Q2 for the ON-beam vs the OFF-beam 2. 
  \textit{Panel b:} Q2 for the OFF-beam 1 vs the OFF-beam 2.
  \textit{Panel c:} Q4a (number of peaks). \textit{Panel d:} Q4b (power of peaks). \textit{Panel e:} Q4e (peak offset). \textit{Panel f:} Q4f (peak offset). For \textit{panels c} to \textit{f} the black lines are the ON-beam difference with the OFF beam 1 and the red lines are the OFF-beam difference. The dashed lines are statistical limits (1, 2, 3$\sigma$) of the difference between all the Q4 values derived using two different Gaussian distributions (each performed 10000 times). We do not see any excess signal in the ON-beam compared to the OFF-beams. In fact, the OFF-beams show a false-positive weak signal at $\eta\sim$4. 
}
  \label{fig:Nondetection}
\end{figure*}

%\textbf{reword below:} 
We performed the post-processing of the LOFAR data exactly as in \citetalias{Turner2021}. A very detailed description of the burst and slow emission post-processing observables can be found in \citetalias{Turner2019} and \citetalias{Turner2021}. A brief summary is given below. The post-processing is performed on the absolute value of the corrected Stokes-V data (as defined in \citetalias{Turner2019}). For the slowly varying emission, we calculate the Q1a time series (e.g. Figure \ref{fig:Q1_all}a-b) and Q1b integrated spectrum (e.g. Figure \ref{fig:Q1_all}c-d) for each beam (ON, OFF1, OFF2, and OFF3). For burst emission, we only use the Q2 and Q4a-f observable quantities. These quantities are designed to find bursty emission (with time scales $<\sim$ 1 minute). The Q2 observable is a time series that is created by high-pass filtering the processed dynamic spectrum and integrating over a large frequency range (e.g. 10 MHz) and time resolution (1 second). This integration over a large frequency range is needed to search for faint emission (see e.g. \citetalias{Turner2021}). Q2 is calculated separately for each beam (ON, OFF1, OFF2, and OFF3). Q2 can be represented by a ``scatter plot'' comparing a pair of beams (e.g. the ON and one of the OFF beams, Figure \ref{fig:Nondetection}a-b). Next, we calculate the Q4a to Q4f statistical measures to provide a statistical view of the entire observation as individual bursts in the Q2 scatter plots are hard to identify. When examining Q4a-f, the ON and one of the OFF time series are compared to each other; for this, we introduce the difference curve Q4f$_{\text{Diff}}$=Q4f(ON)-Q4f(OFF). We then plot this curve against a reference curve computed from 10000 draws of purely Gaussian noise. The post-processing was performed separately over 3 different frequency ranges (15-26 MHz, 15-39 MHz, and 26-29 MHz).

%%%%%%%%%%%%%%%%%%%%%%%%%%%%%%%%%%%%%%
% Q4f L570725 Non-Detection 
%%%%%%%%%%%%%%%%%%%%%%%%%%%%%%%%%%%%%%

%----------------------------------------------------------------------
%----------------------------------------------------------------------
%----------------------------------------------------------------------
\section{Data analysis and results}  \label{sec:dataAnalysis}

We searched for excess signal in the ON-beam both by eye and using the automated search procedure outlined in \citetalias{Turner2021}. To search for a possible detection of burst emission we applied the criteria ($N$1 to $N$12) defined in \citetalias{Turner2019} and \citetalias{Turner2021}. We do not find any slow or burst emissions in the observations. The time series (Q1a) and integrated spectra (Q1b) for all beams can be found in Figure \ref{fig:Q1_all}. The top and bottom panels of Figure \ref{fig:Q1_all} show the time series and integrated spectra, respectively. Unlike in \citetalias{Turner2021}, we do not see a sinusoidal pattern (that may have been caused by imperfect phasing of the array) in the integrated spectrum (Figure \ref{fig:Q1_all}c-d). Next, we examine whether there are any systematic differences in the noise between beams. To do this, we tested for the presence of red noise in the difference time series and integrated spectra (bottom panels Figure \ref{fig:Q1_all}a-d) for all the observations using the time-averaging method (\citealt{Pont2006}) and wavelet technique\footnote{The wavelet technique assumes that the time series is an additive combination of noise with Gaussian white noise and red noise (characterized as a power spectral density proportional to $1/f^{\alpha}$).} (\citealt{Carter2009}) as implemented in the \texttt{EXOMOP} code by \citet{Turner2016b,Turner2017transit}, and found none. Thus any differences in the signals between the ON and OFF beams can be explained purely by Gaussian white noise. The burst statistics were also similar between all ON and OFF beams. An example of the diagnostics for a non-detection can be found in Figure \ref{fig:Nondetection}, based on the L776035 (2020-04-15) observation, analyzed within 15-27 MHz. The ON-OFF and OFF-OFF difference curves are very similar to each other. In summary, we did not detect any slow or bursty signals in our data. In Section \ref{sec:discussion}, we will discuss the implications of these non-detections and our derived upper-limits.

%-------------------------------------------------------------------------------
\section{Discussion} \label{sec:discussion}

We do not detect any bursty or slow emission in the new LOFAR observations. If the signals in \citetalias{Turner2021} were coming from the planet we could have expected to detect them again at the same phases. However, this assumes no variability in the radio signal and that the signals are from the planet. We discuss these possibilities more below. Our non-detection is consistent with the non-detection of bursty emission from five NenuFAR observations \citep{Turner2023} taken simultaneously with the LOFAR observation (see Figure 1 in \citealt{Turner2023}). Most importantly, both NenuFAR and LOFAR don't observe bursty emission on April 15, 2020, which covered the same phase as the \citetalias{Turner2021} bursty tentative detection.

\subsection{Degenerate causes of the non-detection} \label{sec:nondetection}

There are many different degenerate reasons which could explain our non-detection of emission from $\tau~$Boo. We explore the possibilities in greater detail below:
%\begin{enumerate}[(a)]
\begin{enumerate}
\item The first possibility is that the original slow and/or bursty signals seen by \citetalias{Turner2021} were caused by statistical anomalies or unknown unidentified instrumental systematics. In \citetalias{Turner2021}, it was suggested that the OFF beam on the slow emission detection behaved abnormally (the flux on that date was much lower than all the other dates) compared to the rest of the observations. This possibility can be investigated further with the new observations. We find that the OFF beams in the L570725 are not abnormal when compared to the OFF beams in the new 2020 $\tau~$Boo observations (see Figure \ref{fig:Q1b_compare}). This suggests that the original slow emission signal may have not been an unknown instrumental systematic but it doesn't completely it rule out.  

\begin{figure}[!tbh]
    \centering
    \begin{tabular}{c}
         \includegraphics[page=1,width=0.46\textwidth]{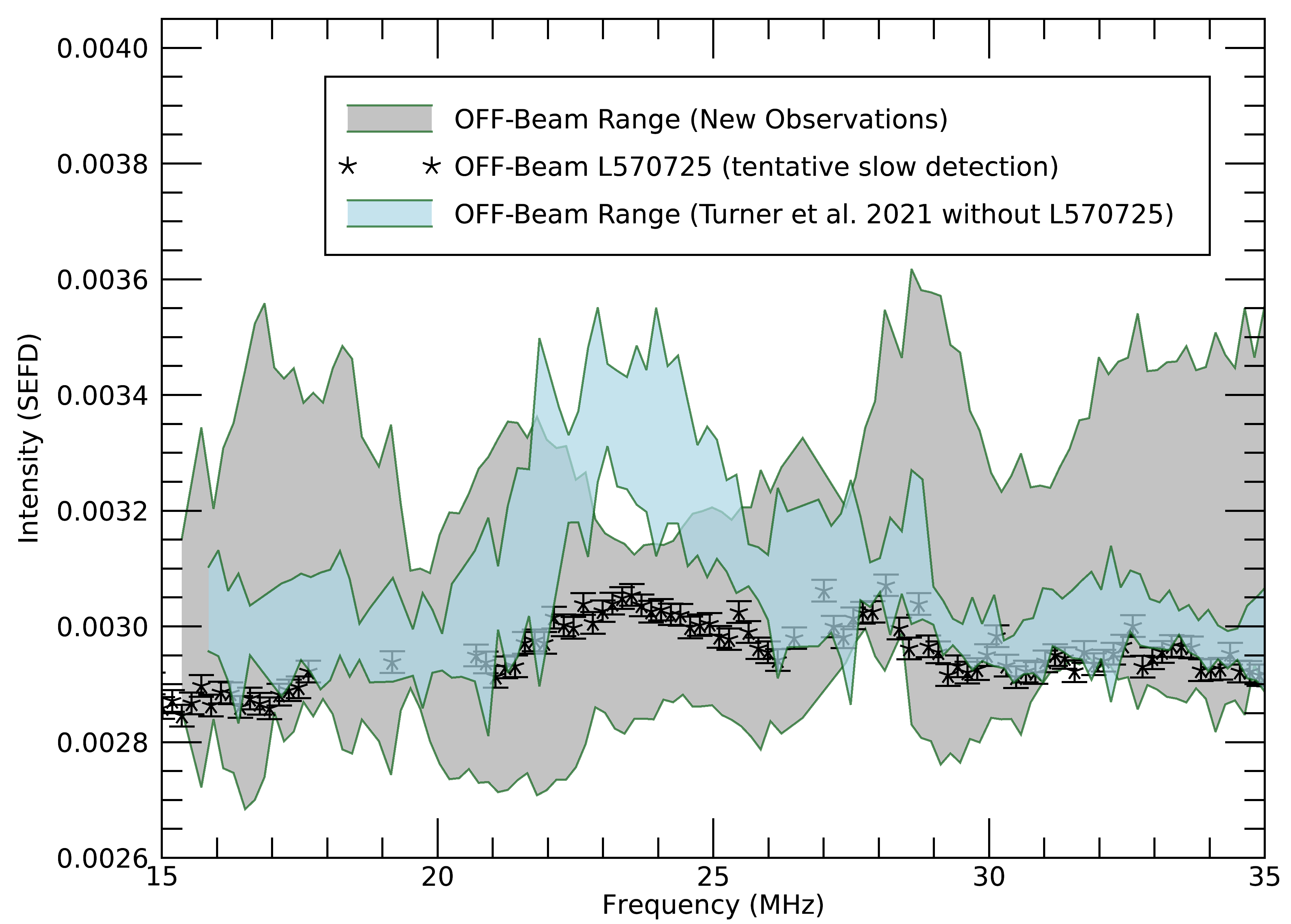} 
    \end{tabular}
    \caption{Comparison of the integrated spectra (Q1b) of the LOFAR $\tau$~Bo\"{o}tis OFF-beams between the new observations in this paper (gray-shaded area) and those presented (light-blue shaded area) in \citet{Turner2021}. We show separately the OFF-beam in the L570725 observation (square black points), where a tentative detection of slow emission was found from \citet{Turner2021}. The integrated spectrum for each date in the current observing campaign can be found in Figure \ref{fig:Q1_all}. While the OFF-beam in L570725 is abnormal when compared to the rest of the \citet{Turner2021} observations, it is consistent with the new observations. }
    \label{fig:Q1b_compare}
\end{figure}
\item The planetary emission from $\tau$ Boo b might be variable. The expected intensity of the planetary radio emission can differ greatly with stellar rotation (e.g. \citealt{Griessmeier05AA,Griessmeier07PSS,Fares2010,Vidotto2012,See2015,Strugarek2022}) and phase of the stellar magnetic cycle (e.g. \citealt{Fares2009,Vidotto2012,See2015,Elekes2023}). The latter point is very much a concern since the host star $\tau~$Boo is known to undergo a rapid magnetic cycle of 120 days (\citealt{Mittag2017,Jeffers2018}). \citet{See2015} showed that the expected radio flux of $\tau~$Boo b can disappear entirely for certain stellar rotation phases and also that it can vary (at optimal stellar longitude) by a factor of $\sim$5 along the stellar magnetic cycle. Recently, \citet{Elekes2023} found using numerical simulations of the $\tau~$Boo system that if the magnetic field of the star were to reverse and become anti-aligned with the magnetic field of the planet, the planetary radio emissions would be significantly reduced and could fall below the detection threshold of LOFAR. Unfortunately, no stellar magnetic field measurements exist during the new LOFAR observations. Therefore, we cannot rule out planetary radio emission variability as the cause of our non-detection. 

\item It is possible that one of the conditions to allow the CMI mechanism to operate is temporarily violated. For example, an extended and evaporating atmosphere may increase the local electron plasma frequency such that it is greater than the local electron cyclotron frequency, $\omega_{p}$ $>$ $\omega_{c}$; (e.g. \citealt{Weber2017,Weber2017pre8,Elekes2023,Griessmeier23PRE9}). 
However, for $\tau~$Boo~b this effect is predicted to not be a problem due to its large mass \citep{Weber2018}. On the other hand, temporal variations could lead to circumstances where the emission does not operate continuously if the planetary magnetic field is as low as estimated by \citet{Reiners2010}. 
Evidence suggests that the observed field strengths for brown dwarfs (\citealt{Kao2016,Kao2018}) cannot be explained by the \citet{Reiners2010} energy flux and magnetic energy density scaling relation. However, it is currently unclear if this contradiction extends to $\tau$~Boo~b and other hot Jupiters. Therefore, more work is 
needed to understand whether the plasma conditions at $\tau$~Boo~b vary such that the conditions for the CMI are not always fulfilled. 

\item 
In contrast to Jupiter's decametric emissions (\citealt{Zarka1998}), $\tau$~Boo~b's radio emission may not exhibit continuous activity. This would suggest that the magnetospheric dynamics of $\tau$~Boo~b could be different from those of Jupiter. For instance, the density of the magnetosphere might vary and at times be temporarily depleted in energetic electrons.
\end{enumerate}

To determine the true cause of the potential radio variability of the $\tau~$Boo system, it is crucial to conduct an extensive follow-up campaign. To rule out instrumental systematics, multi-site observations are recommended. We also advocate for complimentary imaging observations to follow-up on the $\tau~$Boo tentative detection (\citetalias{Turner2021}). A detailed comparison of the advantages and drawbacks between beamformed and imaging observations can be found in Section 6.3 of \citetalias{Turner2021}. It is necessary to monitor the planet's behavior throughout its orbit and the host star's magnetic cycle to distinguish between conflicting factors. Additionally, measurements of the star's magnetic field must be taken during the radio follow-up campaign. Currently, there is an ongoing collaborative campaign utilizing NenuFAR to study $\tau$ Boo radio emission and several telescopes (TBL/Neo-Narval and CHFT/ESPaDOnS) to monitor the magnetic field of the host star. The results of this study will be presented in future work.

\subsection{Upper-limits on the radio flux density}
With our non-detection, we can place an upper limit on the radio emission from the $\tau$~Boo system at the time of the observations. We find a 3$\sigma$ upper limit of 165 mJy from the range 15–39 MHz using the Q1a observable for the slowly varying emission. The noise characteristics of Q1a for the new observations is lower by a factor of 2 than the noise seen in the \citetalias{Turner2021} detection. Using the attenuated Jupiter modeling done in \citet{Turner2019}, we find for the burst emission an upper limit on the flux density that should less than $10^{5}\times$ the peak flux of Jupiter's decametric burst emission ($\sim5\times10^{6}$ Jy; \citealt{Zarka2004}). 

%-------------------------------------------------------------------------------
\section{Conclusions}  \label{sec:conclusion}
In this study, we performed follow-up low-frequency beamformed radio observations of the $\tau$~Bo\"{o}tis ($\tau$~Boo) exoplanetary system with LOFAR. Previous LOFAR observations showed tentative hints of a detection possibly originating from the exoplanetary system (\citealt{Turner2021}). In the new observations, we do not detect any burst or slow emission from $\tau$~Boo (Figures \ref{fig:Q1_all}--\ref{fig:Nondetection}). The current cause of the non-detection is degenerate (see Section \ref{sec:nondetection}) but could be caused by variability in the planetary radio emission. More radio observations (preferably multi-site) are needed that cover the full orbit of the planet multiple times and at different epochs of the stellar magnetic cycle. Near-simultaneous stellar magnetic maps and stellar monitoring are highly encouraged and needed to disentangle the various competing effects. The newly commissioned NenuFAR and LWA-OVRO are scheduled to observe the $\tau$~Boo exoplanetary system and will help investigate the cause of the possible variable emission.

\section*{Acknowledgements} 
J.D. Turner was supported for this work by NASA through the NASA Hubble Fellowship grant $\#$HST-HF2-51495.001-A awarded by the Space Telescope Science Institute, which is operated by the Association of Universities for Research in Astronomy, Incorporated, under NASA contract NAS5-26555.

P. Zarka acknowledges funding from the ERC N$^\circ$ 101020459—Exoradio.

 This work was supported by the ``Programme National de Plan\'{e}tologie'' (PNP) of CNRS/INSU co-funded by CNES and by the ``Programme National de Physique Stellaire'' (PNPS) of CNRS/INSU co-funded by CEA and CNES. 

This research has made use of the Extrasolar Planet Encyclopaedia (exoplanet.eu) maintained by J. Schneider (\citealt{Schneider2011}), the NASA Exoplanet Archive, which is operated by the California Institute of Technology, under contract with the National Aeronautics and Space Administration under the Exoplanet Exploration Program, and NASA's Astrophysics Data System Bibliographic Services. This research has also made use of Aladin sky atlas developed at CDS, Strasbourg Observatory, France (\citealt{Bonnarel2000}; \citealt{Boch2014}). 
%In this paper, all the physical characteristics for the pulsar B0809+74 were taken from the ATNF Pulsar Catalogue \citep{Manchester2005} located at http://www.atnf.csiro.au/research/pulsar/psrcat. 

This paper is based on data obtained with the International LOFAR Telescope (ILT) under project codes LC7$\_$013 and LC13$\_$027. LOFAR (\citealt{vanHaarlem2013}) is the Low Frequency Array designed and constructed by ASTRON. It has observing, data processing, and data storage facilities in several countries, that are owned by various parties (each with their own funding sources), and that are collectively operated by the ILT foundation under a joint scientific policy. The ILT resources have benefited from the following recent major funding sources: CNRS-INSU, Observatoire de Paris and Universit\'{e} d'Orl\'{e}ans, France; BMBF, MIWF-NRW, MPG, Germany; Science Foundation Ireland (SFI), Department of Business, Enterprise and Innovation (DBEI), Ireland; NWO, The Netherlands; The Science and Technology Facilities Council, UK. 

We thank Mickael Coriat, Cyril Tasse, and Vyacheslav Zakharenko for help with the LC13$\_$027 LOFAR proposal.

We acknowledge the use of the Nan\c{c}ay Data Center computing facility (CDN - Centre de Donn\'{e}es de Nan\c{c}ay). The CDN is hosted by the Nan\c{c}ay Radio Observatory in partnership with Observatoire de Paris, Universit\'{e} d'Orl\'{e}ans, OSUC and the CNRS. The CDN is supported by the R\'{e}gion Centre-Val de Loire, d\'{e}partement du Cher.

%\textbf{NenuFAR sentence.}
%\textbf{verify: Nancay institute has changed its name.} 

We thank the ASTRON staff for their help with these observations. 

We thank the anonymous referee for their useful and thoughtful comments.

During the process of writing this paper, Jake D. Turner's young cousin, Evan Turner, passed away unexpectedly. This paper is dedicated to you Evan, keep looking up.

\textbf{Facilities:} \texttt{LOFAR} (\citealt{vanHaarlem2013})

%% Similar to \facility{}, there is the optional \software command to allow 
%% authors a place to specify which programs were used during the creation of 
%% the manusscript. Authors should list each code and include either a
%% citation or url to the code inside ()s when available.\\
\textbf{Software:} \texttt{BOREALIS} (\citealt*{Vasylieva2015}; \citealt*{Turner2017pre8}; \citealt*{Turner2019}); \texttt{IDL Astronomy Users Library} \citep{Landsman1995}; \texttt{Coyote IDL} created by David Fanning and now maintained by Paulo Penteado (JPL).

\bibliographystyle{aasjournal} % style aa.bst
\bibliography{reference.bib} % your references Yourfile.bib

%%%%%%%%%%%%%%%%%%%%%%%%%%%%%%%%%%%%%%%%%%%%%%%%%%%%%%%%%%
%%%%%%%%%%%%%%%%%%%%%%%%%%%%%%%%%%%%%%%%%%%%%%%%%%%%%%%%%%%%
%%%%%%%%%%%%%%%%%%%%%%%%%%%%%%%%%%%%%%%%%%%%%%%%%%%%%%%%%
%%%%%%%%%%%%%%%%%%%%%%%%%%%%%%%%%%%%%%%%%%%%%%%%%%%%%%%%%%%
\begin{appendix}
%\section{Observational setup}\label{app:Summary_obs}
%\edits{Table \ref{tb:obs} gives the LOFAR observation IDs, dates, times, and orbital phases for each of the observations.}

%%%%%%%%%%%%%%%%%%%%%%%%%%%%%%%%%%%%%%%%%%%%%%%%%%%%%%%%%
%%%%%%%%%%%%%%%%%%%%%%%%%%%%%%%%%%%%%%%%%%%%%%%%%%%%%%%%%%%

%\section{Observational setup}\label{app:Summary_obs}
%%%%%%%%%%%%%%%%%
% Date-Time Table 
%%%%%%%%%%%%%%%%%

\end{appendix}
%----------------------------------------------------------------------------

\end{document}